\documentclass[10pt,journal,compsoc]{IEEEtran}
\usepackage[utf8]{inputenc}
\usepackage{tabularx}
\usepackage{graphicx,color}
\usepackage{balance}
\title{Blockchain and Fog Computing for Cyber-Physical Systems: The Case of Smart Industry}

\author{
\IEEEauthorblockN{\textbf{Ouns Bouachir}, \textit{Member, IEEE}, \textbf{Moayad Aloqaily},
\textit{Member, IEEE}, \textbf{Lewis Tseng}, \textit{Member, IEEE}, and \textbf{Azzedine Boukerche}, \textit{Fellow, IEEE}
} 
\IEEEcompsocitemizethanks{

\IEEEcompsocthanksitem O. Bouachir is with College of Technological Innovation, Zayed University, UAE. \protect E-mail: ouns.bouachir@zu.ac.ae

\IEEEcompsocthanksitem M. Aloqaily is with Al Ain University, UAE. \protect E-mail: maloqaily@xanalytics.ca

\IEEEcompsocthanksitem L. Tesng is with Boston College, Boston, MA, USA. \protect E-mail: lewis.tseng@bc.edu

\IEEEcompsocthanksitem A. Boukerche is with University of Ottawa, Ottawa, ON, Canada. \protect E-mail: boukerch@uottawa.ca
}
}

\begin{document}
\maketitle
\begin{abstract}
Blockchain has revolutionized how transactions are conducted by ensuring secure and auditable peer-to-peer coordination. This is due to both the development of decentralization, and the promotion of trust among peers. Blockchain and fog computing are currently being evaluated as potential support for software and a wide spectrum of applications, ranging from banking practices and digital transactions to cyber-physical systems. These systems are designed to work in highly complex, sometimes even adversarial, environments, and to synchronize heterogeneous machines and manufacturing facilities in cyber computational space, and address critical challenges such as computational complexity, security, trust, and data management. Coupling blockchain with fog computing technologies has the potential to identify and overcome these issues. Thus, this paper presents the knowledge of blockchain and fog computing required to improve cyber-physical systems in terms of quality-of-service, data storage, computing and security.
\end{abstract}


\section{Introduction}


Recent advances in electronic and wireless communication have transformed the Internet-of-Things (IoT) through the development of miniature devices that can use and manage the collection and exchange of data. These advantages have enabled the development of small, cost-effective, and low power multi-functional sensing platforms capable of monitoring and communicating diverse information in various sectors, such as vehicular, healthcare, industry and many more. 

The IoT approach is now combined with innovative algorithm-based machine learning (ML) and artificial intelligence (AI) to extract a piece of valuable information that can be useful by other physical devices creating a cyber-physical system (CPS). CPS consists of a physical part and a cyber part connected via a communication network. The physical part mainly consists of sensors and actuators are used to collect data and to perform tasks based on the gathered information. The collected data is then sent, through the network to the cyber part where it is stored and processed using advanced algorithms-based ML and AI to extract valuable insights that can be translated into actions performed by the physical part. This system is highly relevant to the industry, and other fields like medical and healthcare.
CPS is now being applied throughout a variety of industries, including manufacturing (Industry 4.0), logistics, oil/gas, transportation, energy/services, mining, metallurgy and aviation, and several sensors are used to manufacture and form Industrial CPS (ICPS) that can transform how industries function \cite{IIOT}. ICPS can create autonomous self-servicing/healing machines and enhance inventory management through ML. Based on industrial IoT (IIoT), ICPS can collect data about transactions and send it over the network to cloud servers, where it is analyzed and stored so it is available when required. Due to the rapid growth and diversity of devices connected to the IIoT, traditional centralized network architecture must address new service requirements and challenges, and efficiently identify and provide large volumes of data regarding, security, integrity and privacy and other areas.

 Fog/Edge computing has been used to meet these challenges and achieve better quality of service (QoS) and experience (QoE) by conducting data storage and processing operations physically close to the data source in a distributed infrastructure \cite{Fog2}\cite{fog3}. Blockchain is another method that can complement IIoT systems by providing functionality for secure and reliable systems to store and process data \cite{BCIIoT}\cite{BCIoT}. 
 
 This paper proposes creation of an ICPS based on both the blockchain and Fog/Edge approaches to overcome challenges with regard to security, QoS and data storage. It also discusses the advantages and challenges of the proposed ecosystem and potential future research areas. The paper is organized as follows: It first introduces ICPS and explains the challenges in Section \ref{Sec:IIot}, then provides an overview of two paradigm pillars (blockchain and fog) in Section \ref{Sec:BCFog}. The advantages and challenges of combining ICPS with the mentioned paradigms are detailed in Section \ref{sec:comb1}, and Section \ref{sec: comb2} explains the benefits of ICPS-based blockchain and Fog computing and discusses the challenges and related future research. Section \ref{conclusion} concludes the paper.

\section{Industrial Cyber-Physical System (ICPS)}\label{Sec:IIot}
This section introduces ICPS and details the challenges.
\subsection{Overview}
Today, due to several innovations that enhance manufacturing (e.g., sensors, actuators), industrial practices can be significantly improved by monitoring performance quality.  ICPS, based on IIoT that connects machines and industrial entities such as vehicles, power generators and others, enables the collection, exchange, storage and analysis of data to communicate valuable information and insights that allow quick decisions to be made accurately. Thus, it improves the performance and productivity of industrial processes \cite{IIoTFog}\cite{CPS}.

\begin{figure}[h]
    \centering
    \includegraphics[scale=0.47]{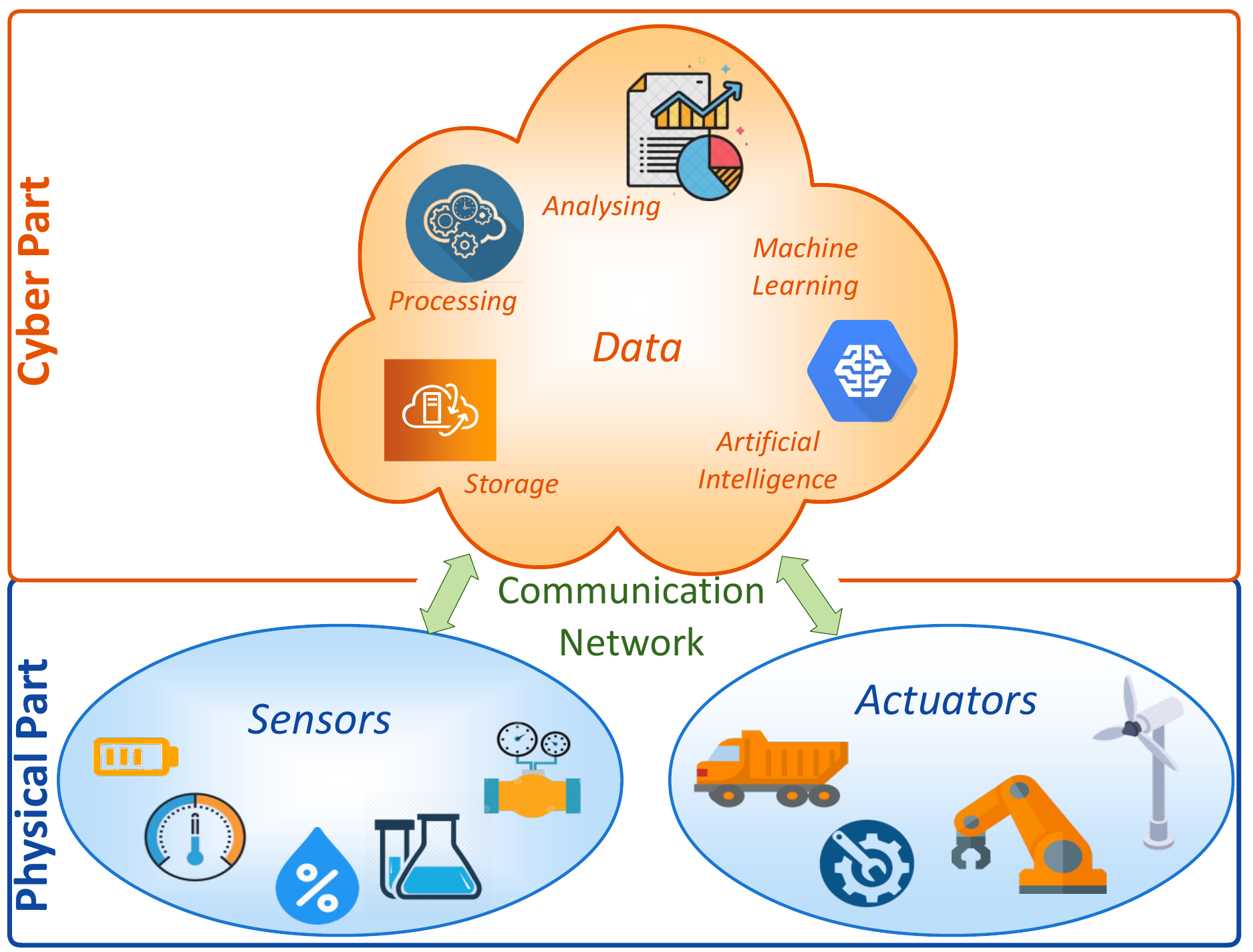}
    \caption{Industrial Cyber-Physical System}
    \label{fig:iiot}
\end{figure}

ICPS combines the power of these smart physical machines with real-time data analysis to achieve higher system efficiency and faster reactions which result in benefits such as saved time and money, better quality control and energy management, asset monitoring, maintenance prediction and decreased waste. Smart industry (Figure \ref{fig:iiot}) is based on a combination of various domains, including, IIoT, robotics, AI, ML and communications, to provide greater efficiency, precision and accuracy. It has also an impact on employee security and health.

\subsection{Challenges}

Typical IoT and IIoT applications are based on wireless sensor network communications that collect and transfer data to a storage center for processing, analysis and storage. These applications, which are deployed in many heterogeneous devices, are continuously producing, exchanging and consuming data through the network, resulting in a significant increase of generated data. The main requirement of IIoT is to capture accurate data in real-time, and then provide rapid and relevant responses that deliver the required performance \cite{IIOT}\cite{IIoTFog}. The main challenges of IIoT are:

\begin{itemize}
\item \textbf{\textit{C1: }Data Storage and manipulation} IIoT devices are equipped with sensors that have minimal computing capabilities and small storage capacity. The collected data is sent, stored and processed in remote clouds. Scaling large numbers of devices and rapidly processing huge volumes of generated data are challenging issues for the current centralized cloud models that are used in the majority of IoT solutions. New solutions require adequate, highly efficient computing power to take advantage of advanced analysis tools and mechanisms that process and store enormous volumes of data and execute extensive applications. Existing centralized cloud paradigms are not suitable to effectively meet many new and future service requirements, such as system availability and efficiency, latency and security in scalable networks.

    \item \textbf{\textit{C2: }Quality of Service (QoS)}
    IIoT applications involve various types of data, including emergency responses, real-time video surveillance, computer vision and self-driving, and all of these have QoS requirements that can change over time, such as delay, throughput and reliability. IIoT must be capable of adapting to these variations, and providing every device with the required service.

\item \textbf{\textit{C3: }Security and Privacy}
In an IIoT application, the network information that is exchanged can be confidential, sensitive, and require protection and provide restricted and controlled access to the data. Most IoT devices are vulnerable, since they have limited security capabilities and can be compromised by hackers relatively easily by accessing stored information or sending incorrect data to clouds. Securing an IIoT system is crucial, and should be guaranteed at both the network and storage system levels.

\end{itemize}

\section{Blockchain and Fog/Edge computing}\label{Sec:BCFog}

Generally speaking, blockchain is a distributed and tamper resistant ledger that does not rely on a centralized authority to establish trust, with a core layer mechanism for decentralized trust management \cite{Salil}. Blockchain mechanisms ensure the system is tamper-proof, since an adversary cannot persuade correct participants to switch to an incorrect branch of the blockchain. This is also how blockchain intuitively establishes decentralized trust.

Blockchain is suitable for IIoT applications as it employs factors such as decentralization, fault-tolerance, immutability and auditability for the operations of cyberphysical systems \cite{p2p}\cite{newOns}. Since it was not designed specifically for IoT applications several challenges need to be addressed, including the energy consumption of consensus mechanisms, communication complexity of verification, storage costs and transaction speed/throughput \cite{Comp2}\cite{Computer1}. These are discussed in more detail in Section \ref{sec:IIoT-BC}.

Fog and Edge computing are more recent data storing architectures that enable moving some sub-processes from centralized to distributed systems\cite{Fog2,fog3}. Since the number of connected devices is continually increasing, the number of exchanges to be handled by Clouds increases accordingly. Moreover, some applications are latency-sensitive; that is, they need fast responses, (particularly data transferring and processing) since delays can lead to serious problems. Thus, cloud network capacity must be increased, and Edge/Fog computing has achieved better QoS and QoE by providing more storage, computing and intelligence resources to devices, as well as various benefits such as the capability to manage huge volumes of data and faster analysis. Table \ref{tab:FogVsCloud} summarizes the differences between these two approaches.

\begin{table}[h]
\caption{Comparison Cloud and Fog computing}\label{tab:FogVsCloud}
\begin{tabularx}{0.5\textwidth}{|l||X|X|X|}
\hline
& Cloud Computing &Fog Computing&Edge Computing\\
\hline
\hline
Geographical Distribution&Centralized&Distributed&Distributed\\
\hline
Location of the Service& Within the Internet&At the edge of the network&At the edge of the network\\
\hline
Distance to end-devices&Through multiple hops (Far)&Single to few hops (Near)&Single hop (very near)\\
\hline
Number of server nodes&Few&Many&Local storage, withing the device\\
\hline
Communication system&IP Network&WLAN, 3G, 4G, IP&None, WLAN\\
\hline
Latency&High&Low&Low\\
\hline
Storage Capacity&High&Very high (Multiple)& Low\\
\hline
\end{tabularx}
\end{table}

Fog and Edge computing seem highly similar due to several key factors, as shown in Table \ref{tab:FogVsCloud}. Both are based on situating data storage, processing and analysis close to data generators (e.g. sensors, motors). With Edge computing, data is either stored and processed locally on the device or sensor itself without being transferred, or on the closest gateway device to the sensors, disregarding the aspect of integration that keeps it discrete. This system can be overwhelmed quickly because of the many small devices, processing and storage capacity. With Fog computing, data is processed by Fog nodes (servers) connected to a LAN or on LAN hardware itself, and the data gathered from several devices can be handled efficiently in real-time. Fog and Edge computing can be combined, as Fog can be considered the standard that defines how to use Edge computing to facilitate and enhance computing and storage operations.

Fog computing involves a set of high-performance physical machines called Fog nodes, linked together in a distributed infrastructure and considered to be a single logical entity. The Fog nodes are located at the edge of the network close to the data source, and they can collect, analyze and process data locally. 
This results in significant traffic mitigation in the core, and faster processing of data services. These nodes can also interact with the Cloud for long term storage when required.

\begin{figure}[]
\centering
\includegraphics[scale=.52]{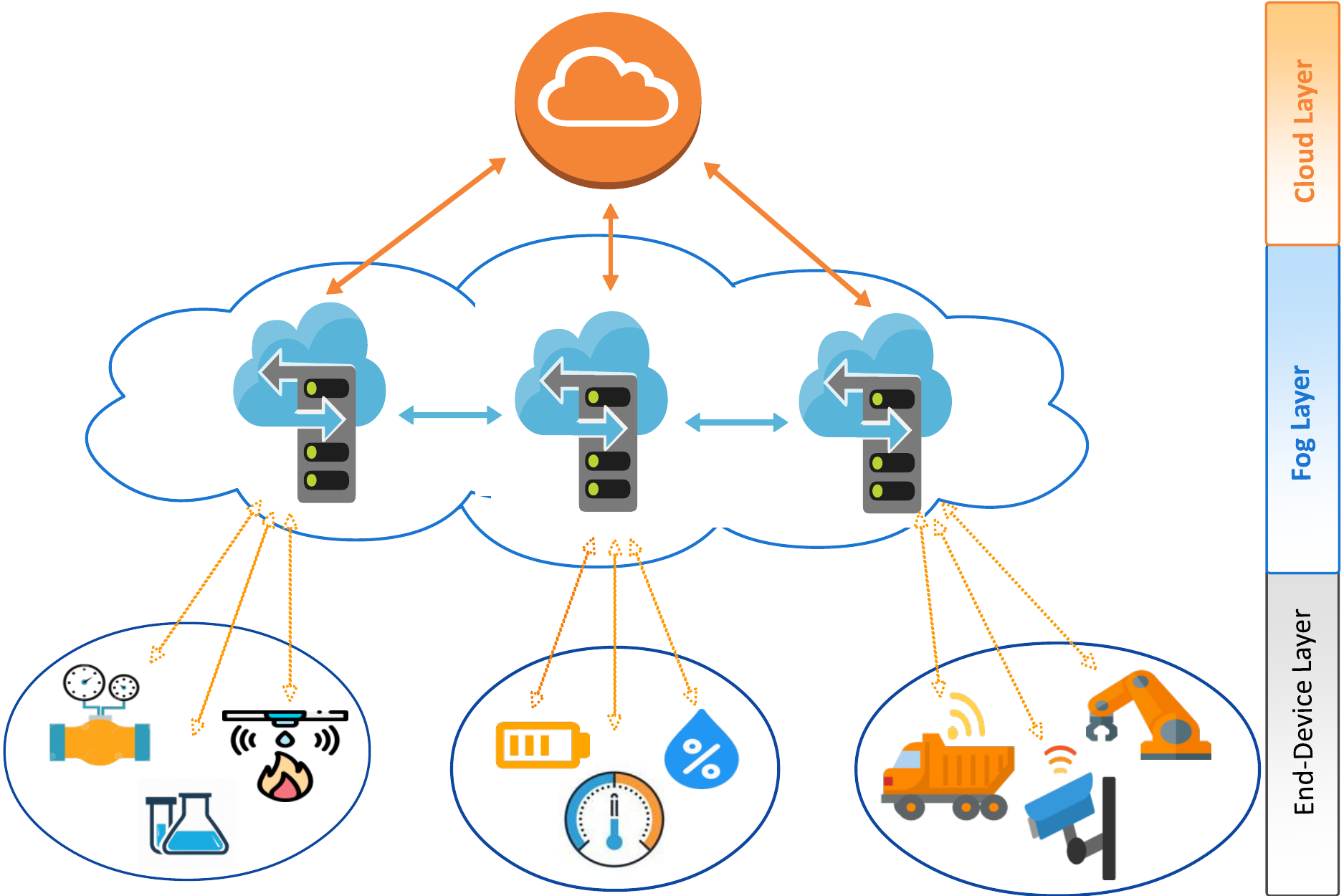}
\caption{Fog computing architecture}\label{fig:fogArchitcture}
\end{figure}

Fog computing can be considered an extension of the Cloud computing paradigm presented in a layered service structure, as shown in Figure \ref{fig:fogArchitcture}. It allows more local real-time monitoring and optimization for IoT applications, while the Cloud provides global optimization and other advanced services.

Fog Architecture consists of three layers:
\begin{itemize}
    \item \textbf{End-Device Layer}: This includes IoT end-user devices that manage data generation. Their main task is to sense surrounding objects and events and forward the data to the upper layers for storage and processing.
\item \textbf{Fog Layer}: As the middle layer, this has a considerable number of Fog nodes (e.g. servers, routers, switches) that are situated at the edge of the network and widely distributed geographically. All these devices are linked together, and they collaborate to store, compute, and exchange sensed data. The connection between fog nodes and end-device layer  is through wireless technologies (e.g. WiFi, 4G, Bluetooth), and each node is connected to the Cloud through an IP internet network. Fog nodes analyse and store received data and send only those deemed valuable to the Cloud server for storage or next processing.

\item \textbf{Cloud Layer}: This includes powerful storage and computing capabilities for substantial computation analysis and permanent storage of major volumes of data. For optimal efficiency, not all computing and storage tasks go through the cloud layer.
\end{itemize}

\section{How to enhance ICPS performance?} \label{sec:comb1}

The best way to overcome 
ICPS challenges and provide high system efficiency is to create an ecosystem that includes modern technologies and paradigms for IIoT systems. Distributed data storage and management are promising solutions to data storage and QoS issues such as Edge and Fog computing mechanisms, and blockchain is also a potential remedy that can be used with these distributed systems to enhance security and privacy. The combination of these paradigms enables ICPS to overcome challenges and achieve higher system performance.

The advantages and challenges of this alliance follow.


\subsection{ICPS based Blockchain}  
\label{sec:IIoT-BC}
Blockchain is a promising concept that will be of benefit to IIoT systems \cite{BCIIoT}\cite{IoTBC5}\cite{Cheng}, since it can provide more secure and reliable data systems (C3).

Deploying blockchain in IIoT systems has several advantages, including:

\begin{itemize}
    \item Security based on cryptographical design;
    \item Immutable data structure;
    \item Transactions and/or data verified by the system;
    \item A total ordering of transactions (or blocks);
    \item Distributed nature (no single point of failure) and P2P interaction, and.
    \item Fault-tolerance due to the use of distributed consensus and gossip-based communication protocol.
\end{itemize}
	A typical blockchain setup requires powerful machines to support heavy crypto primitives and communications, which is not feasible for IIoT. Moreover, IIoT applications utilize different types of devices (e.g. heterogeneity), and if we apply the same blockchain design on heterogeneous devices the reliability and performance properties will vary.
	Traditional Blockchain, especially the consensus protocol, may require prohibitive expensive resources from ICPS small devices. For example, many Proof of Work (PoW) and Proof of Stake (PoS) protocols use cryptographic hash functions that have high energy consumption, and Byzantine Fault Tolerance (BFT) based protocols, e.g., practical BFT, incur high message complexity. Fortunately, several blockchain proposed recently can be used with ICPS such as Plasma Chain or Multi-chains that may ensure better privacy and scalability than the traditional one.
\paragraph*{\textbf{Challenges}}
The main challenges of using blockchain for IIoT is that it is \textit{not} designed to be compatible with heterogeneous small devices with limited computation, communication and storage capabilities \cite{BCIIoT2}. 

\begin{itemize}
    \item \textbf{Resource}: Most IoT devices come with limited battery, computational and storage capabilities to optimise cost and size, whereas most popular blockchains require  powerful machines to support heavy crypto primitives and computation to achieve coordination and provide security. For example, Bitcoin uses PoW, Etherum is adopting PoS and Hyperledger uses PBFT and Raft; all very expensive protocols that require $O(N^2)$  communication complexity, where $N$ is the number of nodes. Additionally, all these protocols require that each node stores an entire copy of the blockchain, which consumes an enormous amount of storage space. Plus, they require public key cryptography to ensure security, another expensive operation. Hence, it is very difficult to apply vanilla blockchain functionality to IIoT applications.

    \item \textbf{Heterogeneity}:
    Most blockchain protocols were designed to be P2P based; that is, each node has approximately the same responsibilities and all have the same power. Though this is useful in most cases, such as financial networks, such a P2P design is \textit{not} adequate for IIoT devices due to cost constraints, as many devices are fragile and should not be required to do anything but push data to others. Moreover, gossip protocol is not the most efficient way to propagate important messages, since IIoT devices could be in sleep mode (to save energy) which would make it less efficient. A final issue is that IIoT scenarios, devices or nodes could move around, thereby making the network dynamic. Typical blockchains perform poorly in dynamic networks.
    
\end{itemize}

\subsection{ICPS based Edge/Fog Computing}

Due to the rapid growth and diversity of devices connected to the IIoT (C1), centralized traditional network architecture is facing new challenges to meet present and future service requirements (C2). Fog/Edge computing 
have been introduced to IoT systems to cope with these challenges, and improve QoS and QoE \cite{IIoTFog}\cite{IIoTFog2}.

IIoT based Fog computing system has several advantages:
\begin{itemize}
\item Adequate computing power to store and process massive volumes of data, and run substantial applications suitable for scalable networks;

\item Achieving higher latency efficiency with many Fog nodes located close to the IIoT devices in different areas of the network, thereby providing fast, high quality localized services with minimal latency to meet the demand of latency-sensitive applications;

\item System capability since tasks are distributed over several high performing Fog level entities;

\item Better network flexibility and availability due to distributed Fog systems in which all Fog nodes are connected and collaborate; this will also maintain normal service if some Fog nodes fail; and,

\item Optimal data traffic and overhead at the core of the network, since the number of re-transmissions in one direction (to the same Cloud server) decreases due to the many receivers located in different areas of the network.

\end{itemize}

\noindent {\textbf{Challenges:}}~
Though Fog/Edge computing provides many advantages for IIoT systems,
there are also challenges to consider when designing a Fog/Edge based IIoT system, such as the heterogeneity of used Fog nodes; indeed, heterogeneous computing nodes can be used to cause a Fog system to manipulate various databases\cite{FogCh1}. This should not be allowed to affect the performance of the applications and communication between Fog nodes and IIoT devices (Fog-Fog and Fog device), which should be optimal, anytime and anywhere. Moreover, securing collected data is another important challenge of a Fog system.
To provide better Fog system performance, three important issues should be taken into consideration:

\begin{enumerate}
    \item How to trust the multiple nodes used as data managing entities?
    \item How to guarantee data integrity in a multi-layered processing system; and,
    \item How to protect the stored data and avoid malicious attacks.
\end{enumerate}
The answers to these questions are based on resolving the following issues:

\begin{itemize}
    \item \textbf{Authentication and Data Integrity}: Fog systems consist of several nodes that communicate with various entities within multiple layers, such as Fog-to-Fog, Fog-to-Cloud and Fog-to-IIoT devices. 
    Fog architecture must ensure secure collaboration and intractability between heterogeneous devices. Authentication can be challenging in such systems, since unlike the unique central authentication server in a Cloud paradigm, the nodes communicate and provide service for various heterogeneous fog devices. Similarly, data integrity is also important since the data is processed by several Fog and Cloud devices, and must be distributed securely without changing it. 
 
   \item \textbf{Malicious Attacks}: Fog nodes are positioned close to IIoT devices where protection and surveillance are relatively weak, and this makes the system vulnerable to diverse malicious attacks that do not occur with Cloud computing, such as denial of services and man-in-the-middle attacks.
   \begin{itemize}
       \item \textit{denial of services (DOS)}: This occurs when an attacker repeatedly requests infinite processing and storage services from Fog nodes, thereby blocking access to the layer. This can be done at various levels, including at the end device where the attacker could spoof the IP address of the IIoT devices and send multiple fake requests. An attacker could also disable communication with the Fog infrastructure by jamming the wireless channel. 

\item \textit{man-in-the-middle}: This type of attack occurs when an intruder interrupts or senses communication between Fog nodes and replaces the Fog devices.
   \end{itemize}
\end{itemize}
Efficient and secure deployment of Fog infrastructure is required for smooth communication between Fog nodes and IIoT devices, and this remains an open issue since existing security strategies are more suitable for centralized systems rather than distributed and open infrastructures like Fog computing. Though managing databases in a distributed manner, such as with Fog systems, solves issues such as delay, it also introduces other potential problems related to authenticating and trusting the data management entities. Blockchain is one of the most interesting solutions for coping with such requirements.

\section{ICPS based Fog/Edge and Blockchain} \label{sec: comb2}
\subsection{Advantages}
To improve IIoT performance and cope with the aforementioned challenges, we propose an ecosystem based on both blockchain and Fog computing that leverages the advantages of each. Deploying blockchain for IIoT networks provides the distribution feature of Edge/Fog units while maintaining security, trust and privacy, and helps solve security and reliability problems related to IIoT patterns. The advantages and challenges of using each paradigm alone with ICPS are  summarized in Table \ref{tab:my_label}, which shows that the main advantages provided by blockchain (i.e. security, privacy and trust) are the principal issues faced by ICPS based Fog/Edge computing, while the main advantages provided by Fog/Edge systems (i.e. storage and computing capabilities) are the challenges with ICPS based blockchain. Thus, combining these two approaches creates a balance by mixing all the advantages to improve system performance.

\begin{table}[h]
 \caption{ICPS based Blockchain and ICPS based fog computing comparison}
    \label{tab:my_label}
    \centering
    \begin{tabularx}{0.5\textwidth}{|l|X|X|}
    \hline
         &\textbf{Advantages} & \textbf{Challenges}  \\
         \hline
          &--Better system performance&--Data storage and manipulation\\
         \textbf{Cyber-physical} &--High autonomy level  & --QoS and scalability\\
       \textbf{Systems} &--Efficient data utilization &--Security, privacy and trust \\
        &  &--Heterogeneity\\
        
        \hline
        \textbf{ICPS based} & --Security&--Scalability\\ 
        \textbf{Blockchain} &--Privacy and trust & --Data storage and manipulation \\
        
        \hline
       
       \textbf{ ICPS based} &--QoS and scalability&--Privacy and trust\\
        \textbf{Fog Computing} & --Data storage and manipulation &--Heterogeneity\\
        
        \hline
    
    \end{tabularx}
   
\end{table}

\begin{figure}[]
\center
\includegraphics[scale=.48]{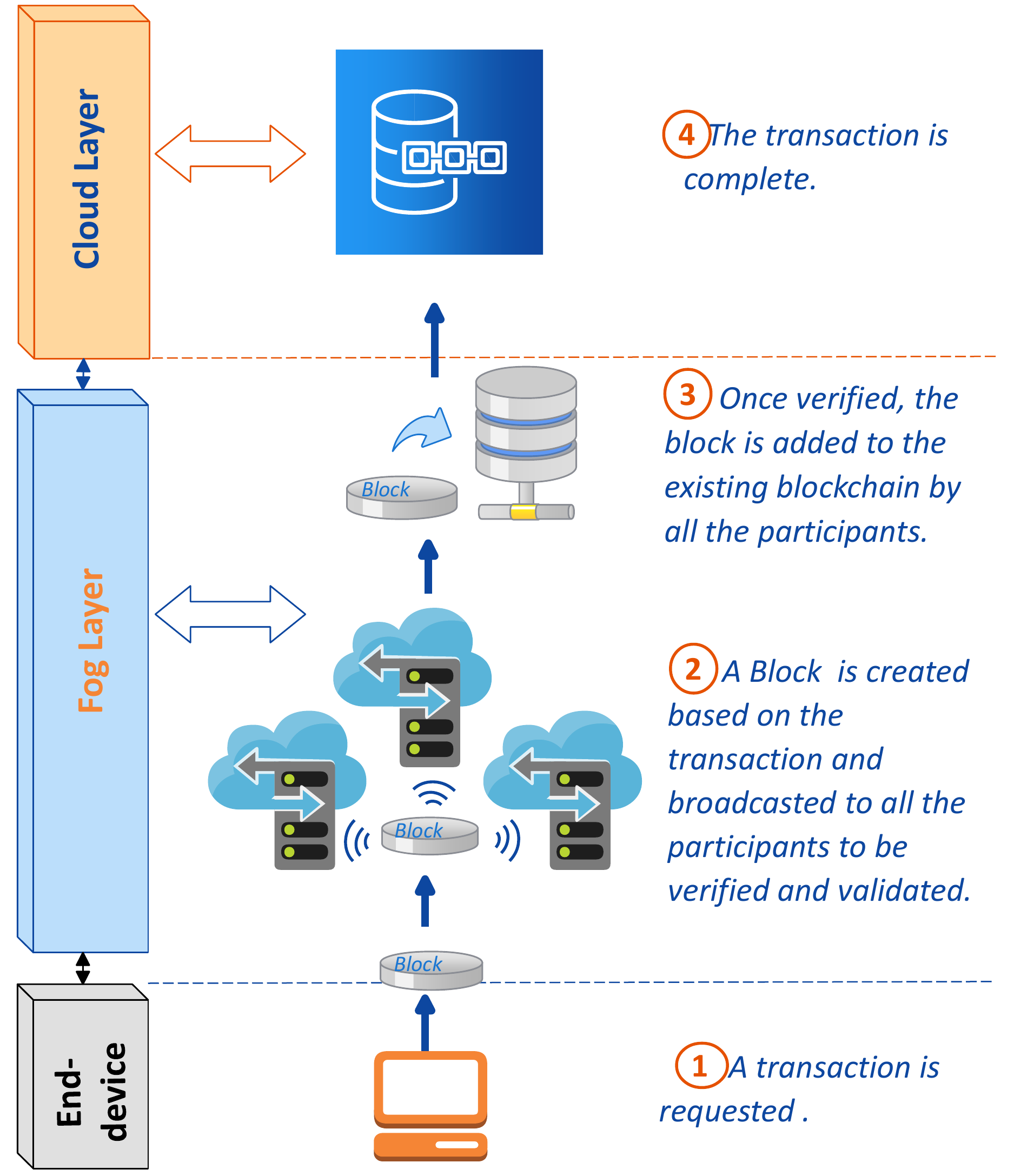} 
\caption{Blockchain for Fog and IIoT}\label{fig:BCIIoTFog}
\end{figure}

The idea of integrating blockchain with IIoT systems raises important questions about the location of the interaction, and the role of Fog computing to facilitate it. The distributed resources of Fog computing are potentially in a location where the mining of IIoT initiatives could occur, which means blockchain can be hosted by the fog and cloud nodes, as shown in Figure \ref{fig:BCIIoTFog}. 

In our design, a transaction is requested by a IIoT device, whose meta-data then be integrated to a block. The block is verified and then added to the integrated blockchain. Finally, the transaction is completed after it is stored on the cloud layer. Security and data integrity come from the verified transaction meta-data stored on blockchain. For example, if the transaction is tampered, then it will not pass the verification since the meta-data (e.g., a hashed value of the transaction) does not match.
Privacy issue such as GDPR compliance can be addressed by using properly designed meta-data. For example, transaction can be removed from the cloud as long as the meta-data does not reveal information, then our system provides GDPR-complied privacy.
The QoS and scalability guarantee from the computing and storage capabilities of the fog nodes.


Ecosystem-based blockchain and Edge/Fog computing with IIoT combines all the advantages discussed in Section \ref{sec:comb1}, and solves many challenges with new benefits, including:

\begin{itemize}
        \item Better management of the distributed feature of blockchain due to the decentralized computing resources of Fog, which provides an improved level of scalability and system availability; and
\item Better system security that ensures privacy, trust, reliability and immunity against malicious attacks. For example, the integrated blockchain can be used for verifying data integrity and mitigate malicious attacks.
\end{itemize}

\subsection{Challenges and future research}
The combination of blockchain and Fog computing addresses many issues inherent in IIoT systems. 
We are developing a benchmarking platform that can be used to evaluate blockchain's performance under varying network constraints. We plan to use the developed tool to test our system.
Below, we list some other interesting unsolved challenges:

\subsubsection{Scalability and storage capacity issues}
Storage capacity and scalability are prominent issues with IIoT and blockchain, since IIoT devices can generate gigabytes (GBs) of data in real-time, which makes this challenge much greater through the utilization of Edge/Fog storage resources. Most current blockchains can only process a limited number of transactions simultaneously, and are not designed to store large volumes of data; attempting to do so increases high latency which is an important barrier and challenge for IIoT.  A detailed description of expected network throughput and network scalability is required to design new techniques allow to simplify real-time processing and storage, including data compression and data lightening.

\subsubsection{Security and data integrity issues}
Data security and protection from cybersecurity attacks are important functions of IIoT systems. In this new industrial evolution, techniques and tools must be developed to counter attacks and provide optimal data security. Though introducing blockchain to IIoT based Fog computing enhances data security and flexibility, it may also impact functions regarding reliability and collected data integrity. Blockchain validates the data generator identity, and ensures the data it is immutable and able to detect any modification of the information. However, the system is limited when data that is already corrupted arrives at the blockchain; it is possible that such corruption is not identified and it remains corrupted. In addition, corrupted data is not only due to malicious attacks, but by other means related the surrounding environment and device failure. 
\subsubsection{Energy Consumption}Solutions like energy aware communication protocols or energy harvesting sensor platforms can complement the proposed platform to decrease the energy consumption in such a system.
\subsubsection{Advanced Algorithms} ICPS based Fog and blockchain is highly innovative automatic system that combines several fields, including robotics, ML and AI able to create smarter cyberphysical system based on context-aware with reasoning capabilities providing autonomous features such as self-monitoring and self-configuration. It requires novel algorithms, applications and interfaces developed using the platform resources to establish new business models.

\subsubsection{Regulation and standards} This ecosystem requires new regulation models. The platform is based on data that could be related to users and can be exchanged between different entities. Thus regulations, standards and social guidelines must be developed to specify how the platform can be used legally and fairly.

\section{Conclusion}\label{conclusion}
This paper discussed the creation of a blockchain and fog computing-based ecosystem in IIoT that can manage and enhance IIoT quality-of-service, data storage and computing and security requirements. Fog/Edge systems transfer computing resources and capacity at the edge of each local network, close to the end-user devices. The main advantages are reduced latency particularly for applications with real-time requirements (e.g. video-surveillance), and increased availability by helping prevent single point of failure issues. IoT-based fog and blockchain systems can handle critical challenges in terms of security, privacy and integrity of collected data. 
Blockchain provides an important level of security for added data. Though IIoT-based blockchain and Fog computing leverages the advantages of both technologies, there are other challenges that will require more research to reveal new opportunities, including the development of algorithms and infrastructures suitable for such highly innovative automatic platforms.

\balance
\bibliographystyle{IEEEtran}
\bibliography{bib}

\begin{thebibliography}{10}
\providecommand{\url}[1]{#1}
\csname url@samestyle\endcsname
\providecommand{\newblock}{\relax}
\providecommand{\bibinfo}[2]{#2}
\providecommand{\BIBentrySTDinterwordspacing}{\spaceskip=0pt\relax}
\providecommand{\BIBentryALTinterwordstretchfactor}{4}
\providecommand{\BIBentryALTinterwordspacing}{\spaceskip=\fontdimen2\font plus
\BIBentryALTinterwordstretchfactor\fontdimen3\font minus
  \fontdimen4\font\relax}
\providecommand{\BIBforeignlanguage}[2]{{%
\expandafter\ifx\csname l@#1\endcsname\relax
\typeout{** WARNING: IEEEtran.bst: No hyphenation pattern has been}%
\typeout{** loaded for the language `#1'. Using the pattern for}%
\typeout{** the default language instead.}%
\else
\language=\csname l@#1\endcsname
\fi
#2}}
\providecommand{\BIBdecl}{\relax}
\BIBdecl

\bibitem{IIOT}
L.~D. {Xu}, W.~{He}, and S.~{Li}, ``Internet of things in industries: A
  survey,'' \emph{IEEE Transactions on Industrial Informatics}, vol.~10, no.~4,
  pp. 2233--2243, Nov 2014.

\bibitem{Fog2}
P.~Hu, S.~Dhelim, H.~Ning, and T.~Qiu, ``Survey on fog computing: Architecture,
  key technologies, applications and open issues,'' \emph{Journal of Network
  and Computer Applications}, vol.~98, 09 2017.

\bibitem{fog3}
I.~A. {Ridhawi}, M.~{Aloqaily}, and A.~{Boukerche}, ``Comparing fog solutions
  for energy efficiency in wireless networks: Challenges and opportunities,''
  \emph{IEEE Wireless Communications}, vol.~26, no.~6, pp. 80--86, December
  2019.

\bibitem{BCIIoT}
N.~{Teslya} and I.~{Ryabchikov}, ``Blockchain-based platform architecture for
  industrial iot,'' in \emph{2017 21st Conference of Open Innovations
  Association (FRUCT)}, Nov 2017, pp. 321--329.

\bibitem{BCIoT}
L.~Tseng, L.~Wong, S.~Otoum, M.~Aloqaily, and J.~B. Othman, ``Blockchain for
  managing heterogeneous internet of things: A perspective architecture,''
  \emph{IEEE Network}, Jan. 2020.

\bibitem{IIoTFog}
R.~Basir, S.~Qaisar, M.~Ali, M.~Aldwairi, M.~I. Ashraf, A.~Mahmood, , and
  M.~Gidlund, ``Fog computing enabling industrial internet of things:
  State-of-the-art and research challenges,'' \emph{sensors}, Nov. 2019.

\bibitem{CPS}
P.~O'Donovan, K.~Bruton, C.~Gallagher, and D.~O'Sullivan, ``A fog computing
  industrial cyber-physical system for embedded low-latency machine learning
  industry 4.0 applications,'' \emph{Manufacturing Letters}, vol.~15, 2018.

\bibitem{Salil}
A.~Dorri, S.~S. Kanhere, R.~Jurdak, and P.~Gauravaram, ``Lsb: A lightweight
  scalable blockchain for iot security and anonymity,'' \emph{Journal of
  Parallel and Distributed Computing}, 2019.

\bibitem{p2p}
F.~Ali, M.~Aloqaily, O.~Alfandi, and O.~Ozkasap, ``Cyberphysical
  blockchain-enabled peer-to-peer energy trading,'' \emph{Computer}, 2020.

\bibitem{newOns}
M.~Aloqaily, A.~Boukerche, O.~Bouachir, F.~Khalid, and S.~Jangsher, ``An energy
  trade framework using smartcontracts: Overview and challenges,'' \emph{IEEE
  Network}, 2020.

\bibitem{Comp2}
A.~{Pal} and K.~{Kant}, ``Using blockchain for provenance and traceability in
  internet of things-integrated food logistics,'' \emph{Computer}, vol.~52,
  no.~12, pp. 94--98, Dec 2019.

\bibitem{Computer1}
H.~{Lei}, C.~{Qiu}, H.~{Yao}, and S.~{Guo}, ``When blockchain-enabled internet
  of things meets cloud computing,'' \emph{Computer}, vol.~52, no.~12, pp.
  16--17, Dec 2019.

\bibitem{IoTBC5}
I.~Makhdoom, M.~Abolhasan, H.~Abbas, and W.~Ni, ``Blockchain's adoption in iot:
  The challenges, and a way forward,'' \emph{Journal of Network and Computer
  Applications}, vol. 125, pp. 251--279, 01 2019.

\bibitem{Cheng}
R.~Singh, A.~Dwivedi, G.~Srivastava, A.~Wiszniewska-Matyszkiel, and X.~Cheng,
  ``A game theoretic analysis of resource mining in blockchain,'' \emph{Cluster
  Computing}, 01 2020.

\bibitem{BCIIoT2}
T.~M. {Fernández-Caramés} and P.~{Fraga-Lamas}, ``A review on the application
  of blockchain to the next generation of cybersecure industry 4.0 smart
  factories,'' \emph{IEEE Access}, vol.~7, pp. 45\,201--45\,218, 2019.

\bibitem{IIoTFog2}
M.~{Aazam}, S.~{Zeadally}, and K.~A. {Harras}, ``Deploying fog computing in
  industrial internet of things and industry 4.0,'' \emph{IEEE Transactions on
  Industrial Informatics}, vol.~14, no.~10, pp. 4674--4682, Oct 2018.

\bibitem{FogCh1}
M.~{Mukherjee}, R.~{Matam}, L.~{Shu}, L.~{Maglaras}, M.~A. {Ferrag},
  N.~{Choudhury}, and V.~{Kumar}, ``Security and privacy in fog computing:
  Challenges,'' \emph{IEEE Access}, vol.~5, pp. 19\,293--19\,304, 2017.

\end{thebibliography}

\vspace{.5cm} 
\textbf{\textit{Ouns Bouachir}} (M'18) is an assistant professor of computer engineering in the College of Technological Innovation, Zayed University, UAE and a researcher in the field of communications and autonomous systems. She has a Ph.D. degree in computer engineering from the French University for Civil Aviation(ENAC), Toulouse, France in 2014.
\vspace{.2cm}

\textbf{\textit{\textbf{Moayad Aloqaily}}} (S'12, M'17) received the Ph.D. degree in Electrical and Computer Engineering from the University of Ottawa, Ottawa, ON, Canada in 2016.  He is the managing director of xAnalytics Inc., Ottawa, ON, Canada. His current research interests include AI and ML, Connected Vehicles, Blockchain Solutions, and Sustainable Energy and Data Management.
\vspace{.2cm}

\textbf{\textit{Lewis Tseng}} is an assistant professor with Boston College, USA. He has a Ph.D. in Computer Science at University of
Illinois at Urbana-Champaign. His research interest includes
distributed computing/systems, blockchain-based systems.

\vspace{.2cm}
\textbf{\textit{Azzedine Boukerche}} (F'15) is a Distinguished University Professor and holds a Canada Research Chair Tier-1 position at the University of Ottawa, Canada. He is a Fellow of the Engineering Institute of Canada, a Fellow of the Canadian Academy of Engineering, and a Fellow of the American Association for the Advancement of Science.

\end{document}